\documentclass[12pt]{article}

\newcommand{\text}{\rm}

\begin{document}

\title{\bf {Vacuum Condensates and Dynamical Mass Generation in
Euclidean Yang-Mills Theories{\footnote{Talk given by S.P. Sorella
at the International Conference on Color Confinement and Hadrons
in Quantum Chromodynamics, Confinement 2003, TITech $\&$ RIKEN,
Tokyo, Japan, July 21-24, 2003.}}}}

\author{D. Dudal\thanks{%
Research Assistant of the Fund for scientific Research-Flanders,
Belgium.},\
 H. Verschelde\thanks{%
david.dudal@ugent.be, henri.verschelde@ugent.be} \\
{\small {\textit{Ghent University }}}\\
{\small {\textit{Department of Mathematical Physics and Astronomy,
Krijgslaan 281-S9, }}}\\
{\small {\textit{B-9000 Gent, Belgium}}} \and V.E.R. Lemes, M.S.
Sarandy, R.F. Sobreiro, S.P. Sorella\thanks{%
vitor@dft.if.uerj.br, sarandy@dft.if.uerj.br,
sobreiro@dft.if.uerj.br,
sorella@uerj.br} \\
{\small {\textit{UERJ - Universidade do Estado do Rio de Janeiro,}}} \\
{\small {\textit{\ Rua S\~{a}o Francisco Xavier 524, 20550-013
Maracan\~{a},
}}} {\small {\textit{Rio de Janeiro, Brazil.}}} \and M. Picariello, A. Vicini\thanks{%
marco.picariello@mi.infn.it, alessandro.vicini@mi.infn.it } \\
{\small {\textit{Universit\'{a} degli Studi di Milano, via Celoria
16,
I-20133, Milano, Italy }}}\\
{\small {\textit{and INFN\ Milano, Italy}}}
\and J.A. Gracey\thanks{jag@amtp.liv.ac.uk } \\
{\small {\textit{Theoretical Physics Division}}} \\
{\small {\textit{Department of Mathematical Sciences}}} \\
{\small {\textit{University of Liverpool }}} \\
{\small {\textit{P.O. Box 147, Liverpool, L69 3BX, United
Kingdom}}}}
\date{}
\maketitle

\begin{abstract}
Vacuum condensates of dimension two and their relevance for the
dynamical mass generation for gluons in Yang-Mills theories are
discussed.
\end{abstract}

\newpage

\section{Introduction}

Vacuum condensates of dimension two in Yang-Mills theories have
witnessed increasing interest in recent years, both from the
theoretical point of view as well as from lattice simulations,
which have provided rather strong evidence for an effective
dynamically generated gluon mass. Here we shall report on our
current work on these condensates. Our aim is that of defining a
renormalizable effective potential and evaluating it in analytic
form. The condensates arise thus as nontrivial solutions of the
gap equation corresponding to the minimization of the effective
potential. The content of this work is as follows. In Sect.2 we
review the gluon condensate $\left\langle A^{2}\right\rangle$ in
the Landau gauge. In Sect.3 the mixed gluon-ghost condensate
$\left\langle \frac{1}{2}A_{\mu }^{\alpha }A_{\mu }^{\alpha }+\xi
\overline{c}^{\alpha }c^{\alpha }\right\rangle $ is analysed in
the Maximal Abelian and Curci-Ferrari gauges. Sect.4 contains a
few remarks on the gauge (in)dependence of the obtained effective
gluon mass and on possible relationships with lattice simulations.
We conclude with a short discussion on the operator $A^{2}$ in
linear covariant $\alpha$-gauges.
\section{The condensate $\left\langle
A^{2}\right\rangle $ in the Landau gauge } The gluon condensate
$\left\langle A^{2}\right\rangle=\left\langle
A^{a}_{\mu}A^{a}_{\mu}\right\rangle$ in the Landau gauge has been
introduced \cite{pb,gz} in order to account for the discrepancy
between the expected perturbative behavior and the lattice results
concerning the two and three point functions in pure Yang-Mills
theories. The lattice estimate for the condensate is \cite{pb} $
\left\langle A^{2}\right\rangle \approx (1.64\;GeV)^{2}$.
\noindent A simple argument shows that $\int d^{4}xA^{2}$ is
invariant under infinitesimal gauge transformations, $\delta
A_{\mu }^{a}=-\left( D_{\mu }\omega \right) ^{a}$, in the Landau
gauge, $\partial A=0$, namely $\delta \int d^{4}xA^{2}=0 $. In the
BRST framework, it turns out that $\int d^{4}xA^{2}$ is BRST
invariant on shell
\begin{equation}
s\int d^{4}xA^{2}=0+{\rm Eqs.{\ }motion}   \label{q1}
\end{equation}
This property ensures that the local operator $A^{2}$ is
multiplicatively renormalizable to all orders of perturbation
theory \cite{dvs}. Its anomalous dimension $\gamma _{\,A^{2}}$ can
be expressed \cite{dvs} as a combination of the gauge beta
function $\beta$ and of the anomalous dimension $\gamma _{A}$ of
the gauge field $A^{a}_{\mu}$, {\it i.e.},
\begin{equation}
\gamma_{A^{2}}=-\left( \frac{\beta(a) }{a}+\gamma _{A}(a)\right)
\;, {\ }{\ }{\ }{\ } a=\frac{g^{2}}{16\pi ^{2}}  \label{q2}
\end{equation}
Concerning the analytic evaluation of $\left\langle
A^{2}\right\rangle$, the two-loop effective potential for $
\left\langle A^{2}\right\rangle $ has been obtained \cite{vkav} in
pure Yang-Mills theory by combining the Local Composite Operators
technique with the Renormalization Group Equations. This has led
to a gap equation whose weak coupling solution gives a
nonvanishing condensate, resulting in an effective gluon mass
$m_{{\rm gluon}}\approx 500MeV$. Recently, the inclusion of
massless quarks has been worked out \cite{Browne:2003uv}. It is
worth mentioning that an effective gluon mass has been reported in
lattice simulations in the Landau gauge \cite{lrg}, yielding $
m_{{\rm gluon}}\approx 600MeV$.
\section{ Other Gauges }
\subsection{The Maximal Abelian Gauge}
The so called Maximal Abelian gauge (MAG) plays an important role
for the dual superconductivity picture for confinement. In the
MAG, the gauge field is decomposed
according to the generators of the Cartan subgroup of the gauge group. For $%
SU(2)$, we have $ A_{\mu }^{a}T^{a}=A_{\mu }T^{3}+A_{\mu }^{\alpha
}T^{\alpha }$ with $\alpha =1,2$. The gauge fixing term in the MAG
is given by
\begin{equation}
\int d^{4}x\left[ \frac{1}{2\xi }F^{\alpha }F^{\alpha
}-\overline{c}^{\alpha }M^{\alpha \beta }c^{\beta }-g^{2}\xi
\left( \overline{c}^{\alpha }\varepsilon ^{\alpha \beta }c^{\beta
}\right) ^{2}\right]  \label{r6}
\end{equation}
where $\xi$ is the gauge parameter and
\begin{equation}
F^{\alpha }=D_{\mu
}^{\alpha \beta }A_{\mu }^{\beta }=\left(
\partial _{\mu }A_{\mu }^{\alpha }+g\varepsilon ^{\alpha \beta
}A_{\mu }A_{\mu }^{\beta }\right)  \label{q3}
\end{equation}
with
\begin{equation}
M^{\alpha \beta }=(D_{\mu }^{\alpha \gamma }D_{\mu }^{\gamma \beta
}+g^{2}\varepsilon ^{\alpha \gamma }\varepsilon ^{\beta \sigma
}A_{\mu }^{\gamma }A_{\mu }^{\sigma }) \label{q4}
\end{equation}
The MAG allows for a residual local $U(1)$ invariance, which has
to be fixed later on. Lattice simulations have shown that the
off-diagonal components $A_{\mu }^{\beta }$ acquire a mass
\cite{as,bc}, reporting $m_{{\rm off-diag}}\approx 1.2 GeV$.
\noindent The operator $A^{2}$ can be generalized to the MAG
\cite{k}. Indeed, the gluon-ghost dimension two operator $O_{{\rm
MAG}}= \left( \frac{1}{2} A_{\mu }^{\alpha }A_{\mu }^{\alpha }+\xi
\overline{c}^{\alpha }c^{\alpha }\right)$ turns out to be BRST
invariant on-shell, namely
\begin{equation}
s\int d^4x \; O_{{\rm MAG}}=0+{\rm Eqs.\;motion} \label{q5}
\end{equation}
The condensate $\left\langle O_{%
{\rm MAG}}\right\rangle $ could thus provide effective masses for
the off-diagonal components. However, at present, very little is
known
about the condensate $\left\langle O_{%
{\rm MAG}}\right\rangle $. Concerning the UV properties of
$O_{{\rm MAG}}$, it has been proven \cite{ew,nos} to be
multiplicatively renormalizable. Its anomalous dimension can be
expressed as \cite{ew,nos}
\begin{equation}
\gamma_{O_{{\rm MAG}}}=-2\left( \frac{\beta(a) }{a}+\gamma_{c_{\rm
diag}}\right)  \label{q6}
\end{equation}
where $\gamma_{c_{\rm diag}}$ is the anomalous dimension of the
diagonal ghost field.
\subsection{The Curci-Ferrari gauge  }
The Curci-Ferrari gauge shares similarities with the MAG,
providing useful information on the gluon-ghost condensate. The
gauge fixing term is
\begin{equation}
\int  d^{4}x  \left( \frac{\left( \partial A^{a}\right) ^{2}}{2\xi } +%
\overline{c}^{a}\partial _{\mu }D_{\mu }c^{a}+\frac{\xi
g}{2}f^{abc}\partial A^{a}\overline{c}^{b}c^{c}\right.
\left. -\frac{\xi g^{2}}{16}f^{abc}\overline{c}^{a}\overline{c}%
^{b}f^{mnc}c^{m}c^{n}\;\right) \label{q7}
\end{equation}
where $a,b,c=1,...,(N^{2}-1)$ for $SU(N)$. For the gluon-ghost
operator\cite{k} in the CF gauge we have
$ O_{{\rm CF}}=\left( \frac{1}{2} A_{\mu }^{a}A_{\mu }^{a}+\xi \overline{c}%
^{a}c^{a}\right)$. The operator $O_{{\rm CF}}$ is renormalizable
to all orders, its anomalous dimension being given by \cite{nos}
\begin{equation}
\gamma_{O_{{\rm CF}}}=-2\left( \gamma_{c}+ \gamma_{gc^2}\right)
\label{q8}
\end{equation}
where $\gamma_{gc^2}$ is the anomalous dimension of the composite
operator corresponding to the BRST variation of the ghost $c$,
{\it i.e.}, $sc=\frac{g}{2}f^{abc}c^bc^c$. Recently, the effective
potential for $O_{{\rm CF}}$ has been obtained \cite{nos1},
leading to a nonvanishing
condensate $\left\langle O_{{\rm CF}%
}\right\rangle $ and to a dynamical gluon mass. Something similar
is expected to happen in the MAG.
\section {Open problems and future perspectives}
Many aspects concerning the gauge condensates of dimensions two
are still open, deserving a better understanding. Two important
questions which arise almost immediately are:

\noindent $\bullet$ 1) are these condensates related to the
effective gluon masses observed in lattice simulations?

\noindent $\bullet$ 2) to what extent are these condensates and
the related gluon masses gauge (in)dependent?

\noindent Concerning the first question, it could be useful to
observe that, to the first approximation, the condensates
$\left\langle A^{2}\right\rangle$,  $\left\langle A^{2}/2 + \xi
\overline{c} c\right\rangle$ modify the euclidean gluon propagator
according to
\begin{equation}
\frac{1}{k^2} \rightarrow \frac{1}{k^2+m^2_{\rm eff}}  \label{q9}
\end{equation}
This suggests that the lattice data could be fitted with a formula
which contains a Yukawa type term. This seems to be the case for
the off-diagonal masses in the MAG \cite{as,bc}. Also, it is worth
remarking that the fitting formula employed in the lattice
simulations for the gluon propagator in the Landau gauge
\cite{lrg}
\begin{equation}
\sum_{a,\mu }\left\langle A_{\mu
}^{a}(k)A_{\mu }^{a}(-k)\right\rangle  =\frac{N}{k^{2}+m_{A}^{2}}\left[ \frac{1}{k^{4}+m_{2}^{4}}+%
\frac{s}{\left( \log (m_{L}^{2}+k^{2})\right) ^{13/22}}\right]
\label{r13}
\end{equation}
contains in fact a Yukawa term. Expression (\ref{r13}) fits well
the lattice data with
\[
m_{A} =0.64GeV,\;\;m_{2}=1.31GeV,\;\;
m_{L}=1.23GeV,\;\;s=0.32,\;\;N=8.1 \]
Concerning the second
question, we look at other gauges allowing for a local
renormalizable operator which could condense, providing thus an
effective gluon mass. That is why we are studying in detail the
local operator $A^2$ in linear covariant $\alpha$-gauges, whose
gauge fixing term is
\begin{equation}
\int  d^{4}x  \left( \frac{1}{2\alpha } \left( \partial A^{a}\right) ^{2}-%
\partial _{\mu }\overline{c}^{a}D_{\mu }c^{a}\;\right) \label{r14}
\end{equation}
Notice that in this gauge the operator $A^2$ is not BRST invariant
on-shell. Nevertheless, we have strong indications of its
multiplicative renormalizability to all orders of perturbation
theory \cite{prep}. There is a simple understanding of this
property. In linear $\alpha$-gauges, due to the shift symmetry of
the antighost, {\it i.e.} $\overline{c} \rightarrow \overline{c} +
{\rm const.}$, the operator $A^2$ cannot mix with the other
dimension two ghost composite operator $\overline{c} c$, which
cannot show up due to the above symmetry. The renormalizability of
$A^2$ is the first step towards the construction of a
renormalizable effective potential in order to study the possible
condensation of $A^2$ and the related dynamical mass generation
\cite{prep}. Although this doesn't yet answer the question of the
gauge (in)dependence, it might increase the number of gauges
displaying such an interesting phenomenon. Finally, it is worth
underlining that an effective gluon mass has been reported in the
Coulomb gauge \cite{jg} as well as in the lattice Laplacian gauge
\cite{lg}.
\section*{Acknowledgments}
We are grateful to the Organizing Committee of the Conference and
to M.N. Chernodub, H. Suganuma, V.I. Zakharov and D. Zwanziger for
valuable and helpful discussions. We are indebted to J. Greensite
for pointing out ref.[14]. The Conselho  Nacional de
Desenvolvimento Cient\'{i}fico e Tecnol\'{o}gico CNPq-Brazil, the
SR2-UERJ, the Coordena{\c{c}}{\~{a}}o de Aperfei{\c{c}}oamento de
Pessoal de N{\'\i}vel Superior CAPES and the MIUR-Italy are
acknowledged for financial support.

\end{document}